\documentclass[12pt]{article}
\usepackage{amssymb,graphicx}

\begin{document}
\title{Spinors for Spinning p-Branes}

\author{Djordje \v{S}ija\v{c}ki\thanks{email: sijacki@phy.bg.ac.yu} \\
Institute of Physics, P.O. Box 57, 11001 Belgrade, Serbia}

\date{}

\maketitle

\begin{abstract}
The group of the $p$-brane world volume preserving diffeomorphism is
considered. The infinite-dimensional spinors of this group are related, by the
nonlinear realization techniques, to the corresponding spinors of its linear
subgroup, that are constructed explicitly. An algebraic construction of the
Virasoro and Neveu-Schwarz-Ramond algebras, based on this infinite-dimensional
spinors and tensors, is demonstrated.
\end{abstract}

\section{Introduction}

The subject of extended objects was initiated in the particle/field theory
framework by the Dirac action for a closed relativistic membrane as the
($2+1$)-dimensional world-volume swept out in spacetime \cite{R1}. It evolved
and become one of the central topics following the Nambu-Goto action for a
closed relativistic string, as the ($1+1$)-dimensional worldsheet area swept
out in spacetime \cite{R2, R3}. An important step was the Polyakov action for
a closed relativistic string, with auxiliary metric \cite{R4}, that enabled
consequent formulations of the Green-Schwarz superstring \cite{R5}, and the
bosonic, and super $p$-branes with manifest spacetime supersymmetry \cite{R6,
  R7}. In this work, we follow the original path of the Nambu-Goto-like
formulation of the bosonic $p$-brane and address the question of the spinors
of the brane world-volume symmetries. For $p =1$, these spinors are well
known, and represent an important ingredient of the spinning string
formulation and the Neveu-Schwarz-Ramond infinite algebras \cite{R8, R9}.

There is a direct connection between the spinors appearing in the $p$-brane
formulation and the world spinors of the Metric-Affine \cite{R10} and
Gauge-Affine \cite{R11} theories of gravity, in a generic non-Riemannian
spacetime of arbitrary torsion and curvature. This is due to a common
geometric and group-theoretic structure of both theories. 

In this work we study the topological and group-theoretical features of the
brane world volume symmetries relevant for the spinor description, we
utilize nonlinear realization techniques to relate these spinors to the ones
of the group of linear transformations, and construct explicitly the latter
ones. Finally, we demonstrate, in the case of spinning string, a
group-theoretical derivation of the Virasoro and Neveu-Schwarz-Ramond algebras,
based on algebraic properties of the corresponding infinite-dimensional
tensorial and spinorial representations.  

\section{$p$-Brane world volume symmetries}

Consider a bosonic $p$-brane embedded in a $D$-dimensional flat
Minkowski spacetime $M^{1,D-1}$. The classical Dirac-Nambu-Goto-like action
for $p$-brane is given by the volume of the world volume swept out by the
extended object in the course of its evolution from some initial to some final
configuration:
\begin{equation}
S = -\frac{1}{\kappa} \int d^{p+1}\xi \ \sqrt{-det \partial_i X^m
  \partial_j X^n \eta_{mn} } \ , 
\end{equation}
where $i$ $=$ $0,1,\dots , p$ labels the coordinates $\xi^i =
(\tau, \sigma_1, \sigma_2, \dots)$ of the brane world volume with metric
$\gamma_{ij}(\xi)$, and $\gamma=\det(\gamma_{ij})$; $m$ $=$ $0,1,\dots , D-1$
labels the target space coordinates $X^{m}(\xi^{i})$ with metric $\eta_{mn}$.
The world volume metric $\gamma_{ij}$ $=$ $\partial_i X^m \partial_j X^n
\eta_{mn}$ is induced from the spacetime metric $\eta_{mn}$.

The Poincar\'e $P(1,D-1)$ group, i.e. its homogeneous Lorentz subgroup
$SO(1,D-1)$, are the physically relevant spacetime symmetries, while the
($p+1$)-dimensional brane world volume is preserved by the homogeneous volume
preserving subgroup $SDiff_{0}(p+1,R)$ of the General Coordinate
Transformation (GCT) group $Diff(p+1,R)$.

The $sdiff_{0}(p+1,R)$ algebra operators, that generate the $SDiff_{0}(p+1,R)$
group, are given as follows, 
\begin{equation}
sdiff_{0}(p+1,R) = \left\{ L_{(n)k}^{i_1i_2\dots i_{n-1}} =
\xi^{i_1}\xi^{i_2}\dots\xi^{i_{n-1}}\frac{\partial}{\partial\xi^{k}} \quad
\vert \quad n = 2, 3, \dots \infty \right\} .
\end{equation}
Preservation of the world volume requires the $L_{(2)}$ operator to be
traceless as achieved by subtracting the dilation operator, i.e. $L_{(2)k}^{i}
= \xi^{i}\frac{\partial}{\partial\xi^{k}} - \frac{1}{p+1}\delta^{i}_{k}
\xi^{j}\frac{\partial}{\partial\xi^{j}}$.   
The $L_{(n)}$, $n
= 2, 3, \dots \infty$, operators are labeled by the $SL(p+1,R)$ subgroup
representations given by the Young tableaux $[\lambda_1 ,\lambda_2 , \dots
,\lambda_{p}]$ with $\lambda_1$ $=$ $2, 3, \dots \infty$, and $\lambda_2 =
\lambda_3 = \dots = \lambda_p = 1$.

The $SDiff_{0}(p+1,R)$ commutation relations read:
\begin{eqnarray}
&&[L_{(m)k}^{i_1 i_2 \dots i_{m-1}},
L_{(n)l}^{j_1 j_2 \dots j_{n-1}}] \\ 
&&= \delta_k^{j_1} L_{(m+n-2)l}^{i_1 i_2 \dots i_{m-1} j_2 j_3 \dots j_{n-1}} 
+\delta_k^{j_2} L_{(m+n-2)l}^{i_1 i_2 \dots i_{m-1}j_1 j_3 \dots j_{n-1}} 
+ \dots
+\delta_k^{j_{n-1}} L_{(m+n-2)l}^{i_1 i_2 \dots i_{m-1} j_1 j_2 \dots j_{n-2}}
\nonumber \\ 
&&- \delta_l^{i_1} L_{(m+n-2)k}^{i_2 i_3 \dots i_{m-1} j_1 j_2 \dots j_{n-1}} 
- \delta_l^{i_2} L_{(m+n-2)k}^{i_1 i_3 \dots i_{m-1} j_1 j_2 \dots j_{n-1}}
- \dots
- \delta_l^{i_{m-1}} L_{(m+n-2)k}^{i_1 i_2 \dots i_{m-2} j_1 j_2 \dots
  j_{n-1}m \nonumber }. 
\end{eqnarray}

The above symmetry considerations are purely classical. In the quantum case,
the corresponding classical symmetry is modified, up to eventual anomalies, in
two ways: (i) the classical group is replaced by its universal covering
group, and (ii) the group is minimally extended by the $U(1)$ group of phase
factors. The corresponding Lie algebra remains unchanged in the first
case, while in the second one, it can have additional central charges.

The feasible ways how to extend the Dirac-Nambu-Goto bosonic $p$-brane action
by the fermionic degrees of freedom are determined by the universal covering
group $\overline{SDiff}_{0}(p+1,R)$ of the $SDiff_{0}(p+1,R)$ group and the
form of its spinorial representations. In the following we address at first
with the topological issues that define the type of the universal covering of
the $SDiff_{0}(p+1,R)$ group, and subsequently, we face the problem of the
$\overline{SDiff}_{0}(p+1,R)$ group spinorial representations construction.

\section{Existence of the double-covering $\overline{SDiff}_{0}(p,R)$}

Let us state first some relevant mathematical results.

Let $g = k + a + n$ be an Iwasawa decomposition of a semisimple
Lie algebra $g$ over $R$. Let $G$ be any connected Lie group with Lie
algebra $g$, and let $K$, $A$, $N$ be the analytic subgroups of $G$ with
Lie algebras $k$, $a$ and $n$ respectively. The mapping
$(k,a,n)\rightarrow kan$, $(k\in K, a\in A, n\in N)$
is an analytic diffeomorphism of the product manifold $K\times A\times N$
onto $G$, and {\it the groups $A$ and $N$ are simply connected.}

Any semisimple Lie group can be decomposed into the product of the {\it
maximal compact subgroup} $K$, an {\it Abelian group} $A$ and a {\it
nilpotent group} $N$. As a result of the above statement, only $K$ is not
guaranteed to be simply-connected. There exists a universal covering
group $\overline{K}$ of $K$, and thus also a universal covering of $G$:
$\overline{G} \simeq \overline{K} \times  A \times N.$

For the group of volume preserving diffeomorphisms, let $Diff(n,R)$ be
the group of all homeomorphisms $f$ of $R^{n}$ such that $f$ and $f^{-1}$
are of class $C^{1}$. Stewart proved the decomposition $Diff(n,R) = GL(n,R)
\times E \times R^{n}$, where the subgroup $H$ is contractible to a
point. In our case the relevant decomposition is $SDiff_{0}(p+1,R) = SL(p+1,R)
\times E \times R^{p+1}$. Thus, as $SO(p+1)$ is the compact subgroup of
$SL(p+1,R)$, one finds that $SO(p+1)$ {\it is a deformation retract of}
$SDiff_{0}(p+1,R)$.

As a result, there exists a universal covering of the Diffeomorphism group
$\overline{SDiff}_{0}(p+1,R) \simeq \overline{SL}(p+1,R) \times H \times
R^{p+1}$.

{\it Summing up, we note that both $SL(p+1,R)$ and $SDiff_{0}(p+1,R)$ 
have double coverings, defined by $\overline{SO}(p+1) \simeq Spin(p+1)$ the
double-coverings of the $SO(p+1)$ maximal compact subgroup}. 

The universal covering group $\overline{G}$ of a given group $G$ is a group
with the same Lie algebra and with a simply-connected group manifold. A finite
dimensional covering, $\overline{SL}(p+1,R)$
i.e. $\overline{SDiff}_{0}(p+1,R)$, exists provided one can embed
$\overline{SL}(p+1,R)$ into a group of finite complex matrices that contain
$Spin(p+1)$ as subgroup. A scan of the Cartan classical algebras points to the
$SL(p+1,C)$ groups as a natural candidate for the $SL(p+1,R)$ groups
covering. However, there is no match of the defining dimensionalities of the
$SL(p+1,C)$ and $Spin(p+1)$ groups for $p \geq 2$,
$$
dim(SL(p+1,C)) = p+1 \quad < \quad 2^{\left[\frac{p}{2}\right] } =
dim(Spin(p+1)) ,
$$
except for $p+1 = 8$. In the $p+1 = 8$ case, one finds that the orthogonal
subgroup of the $SL(8,R)$ and $SL(8,C)$ groups is $SO(8)$ and not
$Spin(8)$. For a detailed account of the $D=4$ case cf. \cite{R12}. Thus, we
conclude that there are no finite-dimensional covering groups of the 
$SL(p+1,R)$, i.e $\overline{SDiff}_{0}(p+1,R)$  groups for any $p \geq 2$. An
explicit construction of all spinorial, unitary and nonunitary
multiplicity-free \cite{R13} and unitary non-multiplicity-free \cite{R14},
$SL(3,R)$ representations shows that they are indeed all defined in
infinite-dimensional spaces. 

The universal (double) covering groups of the group
$\overline{SDiff}_{0}(p+1,R)$ and its $\overline{SL}(p+1,R)$ subgroup are, 
for $p \geq 2$, the groups of infinite complex matrices. All their spinorial
representations are infinite dimensional. In the reduction of this
representations w.r.t. subgroups $Spin(p+1)$, with a trivial metric tensor
$\delta$, or $Spin(1,p)$, with a Minkowski-like metric tensor $\eta$, one has
representations of unbounded spin values.

\section{The deunitarizing automorphism.}

The unitarity properties, that ensure correct physical description
of the relevant representations of the $\overline{SDiff}_{0}(p+1,R)$ and
$\overline{SL}(p+1,R)$ groups on quantum states and fields, can be achieved by
making use of the unitary (irreducible) representations construction of these
groups and the so called "deunitarizing" automorphism of the
$\overline{SL}(n,R)$ group. This procedure ensures that in the Special
Relativity limit (Lorentz invariance) all physical objects have the usual
properties (i.g. boosted electron and/or quark retain their Poincar\'e
properties). 

The commutation relations of the $\overline{SL}(p+1,R)$ generators
\begin{equation}
Q_{jk} = i \eta_{jl}L_{(2)k}^{l}, \quad j,k,l =  0, 1, \dots , p, \quad
\eta_{jl} = diag(+1, -1, \dots , -1), 
\end{equation}
are 
\begin{equation}
[Q_{ij},Q_{kl}] = i(\eta_{jk}Q_{il} - \eta_{il}Q_{kj}) ,
\end{equation}

The important subalgebras
are as follows.

{\bf (i)} $so(1,p)$: The $M_{ij} = Q_{[ij]}$  operators generate the
Lorentz-like subgroup $\overline{SO}(1,p) \simeq Spin(1,p)$ with $J_{mn} = 
M_{mn}$ (angular momentum) and $K_{m} = M_{0m}$ (the boosts) $m,n = 1,2,\dots
,p$. 

{\bf (ii)} $so(p+1)$: The $R_{\hat i \hat j}$ operators, $\hat i , \hat j = 1,
2, \dots p+1$, i.e. $J_{mn}$ and $N_{m} = Q_{\{0m\}}$ operators generate the
maximal compact subgroup $\overline{SO}(p+1) \simeq Spin(p+1)$.

{\bf (iii)} $sl(p)$: The $J_{mn}$ and $T_{mn} = Q_{\{mn\}}$ operators
generate the subgroup $\overline{SL}(p,R)$ - an analog of the "little" group
of the massive particle states in Poincar\'e theory.

{\it The $\overline{SL}(p+1,R)$ commutation relations are invariant under
the ``deunitarizing'' automorphism} (originally introduced for the $p=3$
case \cite{R12},
\begin{eqnarray*}
 &J^{\prime}_{mn} = J_{mn}\ ,\quad K^{\prime}_{m} = iN_{m}\ ,\quad
N_{m}^{\prime} = iK_{m}\ ,\\
 &T^{\prime}_{mn} = T_{mn}\ ,\quad
T^{\prime}_{00} = T_{00} \ (= Q_{00})\ , \\
\end{eqnarray*}
so that $(J_{mn},\ iK_{m})$ generate the new compact
$\overline{SO}(p+1)^\prime$ and $(J_{mn},\ iN_{m})$ generate
$\overline{SO}(1,p)^\prime$.

The above deunitarizing automorphism generalizes to the arbitrary signature
case. Let $\overline{SL}(n,R)$ group act on $R^{r,s}$, $r+s=n$ with metric
$\eta = diaq(+1,\dots +1, -1,\dots -1)$ having $r$ times $+1$ and $s$ times
$-1$ on the diagonal. The group generators $Q_{ij}$ split accordingly to
$Q_{ab}$, $Q_{mn}$, $Q_{am}$, and  $Q_{ma}$, where $a,b = 1,2\dots r$, $m,n = 
1,2,\dots s$. The deunitarizing automorphism, that leaves the $sl(n,R)$
algebra invariant, is given as follows,    
\begin{equation}
Q^{\prime}_{ab} = Q_{ab}\ ,\quad Q^{\prime}_{mn} = Q_{ab}\ ,\quad
Q^{\prime}_{am} = i Q_{am}\ ,\quad  Q^{\prime}_{ma} = -i Q_{ma}
\end{equation}

The construction of physically relevant representations is achieved through a
two step procedure: (1) One constructs, utilizing the appropriate mathematical
theorems and methods, the unitary irreducible spinorial, as well as tensorial,
representations of the $\overline{SDiff}_{0}(p+1,R)$ and
$\overline{SL}(p+1,R)$ groups in the basis of the maximal compact $Spin(p+1)$
subgroup representations, and (2) One converts these representations, by
making use of the deunitarizing automorphism, to representations that are
finite and nonunitary for the physical $Spin(1,p)$ subgroup.

\section{Nonlinear $\overline{SDiff}_{0}(p+1,R)$ representations}

The GCT group $SDiff_{0}(p+1,R)$ is an infinite parameter Lie group with the
corresponding infinite algebra that acts linearly, e.g. as infinite matrices,
on an infinite dimensional vector space. However, its defining representation
is given by the group of volume preserving nonlinear transformations of the
$R^{p+1}$ spacetime. The $SDiff_{0}(p+1,R)$ group being nonlinearly realized
over its $SL(p+1,R)$ subgroup. 

The defining representation of the $\overline{SDiff}_{0}(p+1,R)$ universal
(i.e. double) covering group, as well as of its $\overline{SL}(p+1,R)$
subgroup, is given, as demonstrated above, by the infinite dimensional
matrices. In other words, there are no group of finite complex matrices that
is isomorphic to $\overline{SDiff}_{0}(p+1,R)$.

Let us consider now the spinorial representations of the
$\overline{SDiff}_{0}(p+1,R)$ group. There are genuine linear spinorial
representations of the $\overline{SDiff}_{0}(p+1,R)$ group that are infinite
dimensional. Moreover, all of its infinitely many Lie algebra generators are
likewise represented linearly by infinite matrices. Besides, there are two
distinct classes of $\overline{SDiff}_{0}(p+1,R)$ nonlinear spinorial
realizations characterized by: 

(i) $\overline{SDiff}_{0}(p+1,R)$ group is nonlinearly realized over its
maximal linear subgroup $\overline{SL}(p+1,R)$; $\overline{SL}(p+1,R)$ and
$Spin(1,p)$ are represented linearly, 

(ii) both $\overline{SDiff}_{0}(p+1,R)$ and its $\overline{SL}(p+1,R)$
subgroup are realized nonlinearly over the orthogonal subgroup $Spin(1,p)$.

We recall now a few basic notions from the nonlinear representations theory
\cite{R15, R16} and set up required notation. Let $G$ be an $n_G$ parameter
Lie group, and let $H$ be an $n_H$ parameter subgroup of $G$. Let $\cal M$ be
a real analytic manifold of dimension $d$. The mappings $R$ from $g \times\cal
M$ into $\cal M$ form a representation of $G$ if, for each $g \in G$, $p \in
\cal M$, there is an element $R(g)[p] \in \cal M$ such that (i) $R:\ (g,p)
\rightarrow R(g)[p]$ is analytic, (ii) $R(e)[p] = p$, for all $p \in \cal M$,
$e$ is the identity in $G$, and (iii) $R(g_1)R(g_2)[p] = R(g_1 g_2)[p]$, for
all $g_1 , g_2 \in G$, all $p \in \cal M$.  

At each point $p \in \cal M$, local coordinates can be introduced by mapping
an open neighborhood of $p$ into an open neighborhood of $R^d$. Let $q$
denotes the coordinates of a general point $p \in \cal M$, and let $\alpha$ be
the group parameters of an element $g \in G$ in a neighborhood of $e$. Then
$R(g)[p]$ can be expressed as an analytic function $r(q,\alpha )$ of both $q$
and $\alpha$, which is in general nonlinear. 

An equivalence of two representations is naturally expressed through an
independence of the choice of coordinates. Usually, there exists a special
point, base point, on $\cal M$ which must be represented by the origin $q_0$
in all coordinates. Thus, one defines a concept of local equivalence. Two
representations $R_1 (g)$ and $R_2 (g)$ are locally equivalent if there exists
an (in general nonlinear) operator $S$ from $R^n \rightarrow R^n$ such that
(i) $S:\ q \rightarrow S[q]$ is analytic and has an analytic inverse at $q_0$,
(ii) $S[R_1 (g)] [q] = R_2 (g) S[q]$, for all $g \in G$ in a suitable
neighborhood of the identity, and all $q$ in a neighborhood of $q_0$, and
(iii) $S[q_0 ] = q_0$. Representation is said linearizable if it is locally
equivalent to e linear representation. 

Let $H$ be a subgroup of $G$ such that for each $h \in H$, $R(h)[q_0 ] = q_0$,
i.e. let $H$ be the isotropy subgroup of the origin $q_0$. Now, it turns out
that a restriction $R(h)$, $h \in H$ of the representation $R(g)$ is locally
equivalent to a linear representation. In the expansion $R(g)[q]$, $g=h \in H$
in power series $R(h) = D(h)q + O(q^2 )$, one finds a linear representation
$D(h)$ of $H$. The change of coordinates defined by $S:\ q \rightarrow \bar q
= S[q] = \int_H dh D^{-1}(h) R(h)[q]$, where $dh$ is the right invariant
measure on $H$, establishes a local equivalence between $D(h)$ and the
restriction of $R(g)$ to $H$, i.e. $R(h)[\bar q] = D(h)\bar q$. 

An arbitrary element $g$ in $G$ can be written as $g = ch$, where $h$ belongs
to $H$ and $c$ belongs to the left coset space $C = G/H$. Furthermore, an
arbitrary point $q$ of the orbit can be written as $q = R(g)[q_0 ] =
R(c)R(h)[q_0 ] = R(c)[q_0 ]$. Thus, the elements of the orbit are in
one-to-one correspondence with the elements of the coset space $G/H$. They
form a homogeneous space on which $G$ can be represented. 

An action of an arbitrary element $g_1$ on $c$ is as follows $g_1 c = c_1 h_1
c = c' h'$. The parameters of the group element $h'$ depend both on the group
element $g_1$ and on $c$, i.e. $h' = h'(c,g_1 )$. The transformation $h
\rightarrow h'$ is in general nonlinear, and it becomes linear when $g_1$ is
restricted to $H$. 

Let us choose the generators $X_a$, $a = 1, 2, \dots , n_H$ of $H$ and the
remaining generators $Y_b$, $b = 1, 2, \dots , n_G - n_H$ of $G$ such that
they form together a complete set of generators of $G$ that is orthonormal
with respect to the Cartan inner product. In some neighborhood of the identity
of $G$, every element $g \in G$ can be decomposed uniquely as follows
\begin{equation}
g = ch = e^{-i \zeta \cdot Y}\ e^{-i \omega \cdot X}, \quad \zeta \cdot Y =
\zeta^b Y_b ,\quad\omega\cdot X = \omega^a X_a ,\quad \zeta^b ,\omega^a \in R .
\end{equation}
The $\zeta^b$ and $\omega^a$ parameters form a real $n_G$-component vector
$(\zeta , \omega)$. Now, owing to the fat that $H$ leaves the origin $q_0$
fixed, the orbit $\cal N$ of $q_0$ under $G$ separates the $G/H$ cosets
defined by $L_\zeta = e^{-i \zeta\cdot Y}$. One has 
$$ 
R(g)[q_0 ] = R(e^{-i \zeta\cdot Y}) R(e^{-i \omega\cdot X})[q_0] = e^{-i
  \zeta\cdot R(Y)}[q_0 ] ,
$$
and the dimension of the orbit $\cal N$ is given by the number of $\zeta^b$
parameters, i.e. it is equal to $n_G - n_H$. The simplest choice is to
represent the orbit elements by $L_\zeta$. We split now the manifold $\cal M$
into $\cal N$ and its orthogonal complement $\cal V$, which is $d - (n_G -
n_H)$ dimensional, i.e. $\cal M = \cal N + \cal V$. Finally, for the
coordinates of $\cal M$ we write $q = (L_\zeta , \psi)$, $L_\zeta \in \cal N$,
$\psi \in \cal V$. According to the linearization procedure, we can chose the
coordinates $(L_\zeta , \psi)$ so that $H$ acts linearly, and in particular
the coordinates $\psi$ span a space of a linear representation $D(h)$ of $H$. 

Owing to $g_1 c = c' h' = c' h(c,g_1 )$, and $c = L_\zeta$, one finds for
$L_\zeta$ the following transformation law, 
\begin{equation}
g:\ L_\zeta \rightarrow L_{\zeta{}'} = g L_\zeta h^{-1}(\zeta ,g),\quad g \in
G, h \in H ,
\end{equation}
while $\psi$ transforms according to 
\begin{equation}
g:\ \psi \rightarrow \psi{}' = D(h(\zeta ,g)) \psi = D(L^{-1}_{\zeta{}'} g
L_\zeta ) \psi = D(e^{-i \omega (\zeta ,g) \cdot X}) \psi .
\end{equation}
When $g = h$, 
$$
L_{\zeta{}'} = h L_\zeta h^{-1} = D^{(\zeta )}(h) L_\zeta , \quad h \in H
$$
where $D^{(\zeta )}$ is a linear representation of $H$ in the $\zeta^b$ space,
while 
$$
\psi{}' = D(h) \psi = D(e^{-i \omega\cdot X}) \psi .
$$ 
For a linear representation $D(g)$, $g \in G$, one has
$$
D(L_\zeta ) \rightarrow D(L_{\zeta{}'}) = D(g L_\zeta h^{-1}(\zeta ,g)) = D(g)
D(L_\zeta ) D(h^{-1}(\zeta ,g)).
$$
Let $\Psi$ be a basis of this linear representation, i.e. $\Psi{}' = D(g)
\Psi$, $g \in G$. By defining 
\begin{equation}
\psi = D(L^{-1}_\zeta ) \Psi ,
\end{equation}
one relates the linear and nonlinear representations, i.e. one project the
linear representation into the corresponding nonlinear one. Indeed, one has 
\begin{equation}
\psi \rightarrow \psi{}' = D(h) \psi , \quad h = h(\zeta ,g) =
L^{-1}_{\zeta{}'} g L_\zeta \in H
\end{equation}
Moreover, one can express the basis $\Psi$ of a linear representation $D(g)$
in terms of the corresponding basis $\psi$ of its nonlinear representation
$D(h(\zeta ,g))$ as follows
\begin{equation}
\Psi = D(L_\zeta ) \psi .
\end{equation}

\subsection{Nonlinear representations over $\overline{SL}(p+1,R)$}

Let us consider the case where $\overline{SDiff}_{0}(p+1,R)$ group is
nonlinearly realized over its maximal linear subgroup
$\overline{SL}(p+1,R)$. This is a natural extension of $SDiff_{0}(p+1,R)$
being linearly realized over $SL(p+1,R)$. 

As stated above, $\overline{SDiff}_{0}(p+1,R) = \overline{SL}(p+1,R) \times E
\times R^{p+1}$, and thus we have now $g \in G = \overline{SDiff}_{0}(p+1,R)$,
$h \in H = \overline{SL}(p+1,R)$, and  $c = L_\zeta \in G/H = E \times
R^{p+1}$.

Let $\psi$ transforms w.r.t. a spinorial representation of the
$\overline{SL}(p+1,R)$ group, i.e.
\begin{equation}
\psi{}'_A = \left( D_{\overline{SL}(p+1,R)}(h)  \right)^B_A
\psi_B , \quad h \in \overline{SL}(p+1,R)\  A, B = 1, 2, \dots \infty
\end{equation}
where the index that enumerates the components of $\psi$ runs over an infinite
range due to the fact that the spinorial representations of the
$\overline{SL}(p+1,R)$ group are for $p+1 \geq 3$ necessarily infinite
dimensional. The $\overline{SDiff}_{0}(p+1,R)$ spinor $\Psi$ transforms as
follows
\begin{equation}
\Psi{}'_{\tilde A} = \left( D_{\overline{SDiff}_0 (p+1,R)}(g)  \right)^{\tilde
  B}_{\tilde A} \Psi_{\tilde B} , \quad g \in 
\overline{SDiff}_0 (p+1,R) ,\  \tilde A , \tilde B = 1, 2, \dots \infty 
\end{equation}
The $D_{\overline{Diff}_0(p+1,R)}$ representations can be reduced to direct
sum of infinite dimensional $D_{\overline{SL}(p+1,R)}$ representations. We
consider here those representations of $\overline{Diff}_0(D,R)$ that are
nonlinearly realized over the maximal linear subgroup $\overline{SL}(D,R)$.

Provided the relevant $D_{\overline{SL}(p+1,R)}$ spinorial representations are
known, one can first define the corresponding spinors, $\psi_{A}$, and than
make use of the infinite-component pseudo-frames
\begin{equation}
E^{A}_{\tilde A} = \left( D(L_\zeta ) \right)^{A}_{\tilde A} 
\end{equation}
to achieve the required linear-to-nonlinear mapping \cite{R17}
\begin{equation}
\Psi_{\tilde{A}} = E^{A}_{\tilde{A}}(x) \Psi_{A}, \quad
E^{A}_{\tilde{A}} \sim \overline{Diff}_{0}(p+1,R)/\overline{SL}(p+1,R)
\end{equation}
The pseudo-frames $E^{A}_{\tilde{A}}$ infinitesimal transformations are given
by 
\begin{equation}
\delta_{\overline{SL}(p+1,R)} E^{A}_{\tilde{A}} = i \epsilon^{i}_{j}
\{Q_{i}^{j}\}^{A}_{B} E^{B}_{\tilde{A}}
\end{equation}
where  $\epsilon^{i}_{j}$ and $Q_{i}^{j}$ are the group parameters and
generators of $\overline{SL}(p+1,R)$, respectively.

The above outlined construction allows one to define a 
$\overline{Diff}(p+1,R)$ covariant Dirac-like wave equation for the
corresponding spinor $\Psi$ provided a Dirac-like wave equation
for the $\overline{SL}(p+1,R)$ group is known. In other words, one can
lift up an $\overline{SL}(p+1,R)$ covariant equation of the form
\begin{equation}
\big( i \big(\Gamma^{k}_{\overline{SL}(p+1)}\big)^{B}_{A}
\partial_{k} - m \big) \psi_{B} = 0, \quad k = 0, 1, \dots ,p  \nonumber
\end{equation}
to a $\overline{Diff}(p+1,R)$ covariant equation
\begin{equation}
\big( i E^{A}_{\tilde{A}}
\big(\Gamma^{k}_{\overline{SL}(p+1)}\big)^{B}_{A}E^{\tilde{B}}_{B}
\partial_{k} - m \big) \Psi_{\tilde{B}} = 0, \quad k = 0, 1, \dots , p
\end{equation}
where the former equation exists provided a spinorial
$\overline{SL}(p+1,R)$ representation for $\psi$ is given, such that the
corresponding representation Hilbert space is invariant
w.r.t. $\Gamma^{i}_{(\overline{SL}(p+1)}$  action. The crucial step towards a 
Dirac-like GCT spinor equation is a construction of the vector
operator $\Gamma^{i}_{\overline{SL}(p+1)}$ in the space of
$\overline{SL}(p+1,R)$ spinorial representations. We have recently presented
an explicite construction of the Diffeomorphism covariant Dirac-like equation
in the $p+1 = 3$ dimensional case \cite{R18}.

\subsection{Nonlinear representations over $Spin(1,p)$}

Let us consider now the case where $\overline{SDiff}_{0}(p+1,R)$ group is
nonlinearly realized over its maximal compact subgroup $Spin(p+1)$ or over the
related, physically more interesting, Lorentz-like group $Spin(1,p)$. 

The relevant group decompositions are: $\overline{SDiff}_{0}(p+1,R) =
\overline{SL}(p+1,R) \times E \times R^{p+1}$, and the Iwasawa decomposition
$\overline{SL}(p+1,R) = Spin(1,p) \times A_{p+1} \times N_{p+1}$, where
$A_{p+1}$ and $N_{p+1}$ are the groups of $(p+1)\times (p=1)$ Abelian and
nilpotent matrices, respectively. Therefore,  $g \in G =
\overline{SDiff}_{0}(p+1,R)$, $h \in H = Spin(1,p)$, and  $c = L_\zeta \in G/H
= E \times R^{p+1} \times A_{p+1} \times N_{p+1}$.

Here, $\psi$ transforms w.r.t. a spinorial representation of the
$Spin(1,p)$ group, i.e.
$$
\psi{}'_{\alpha} = \left( D_{Spin(1,p)}(h)  \right)^{\beta}_{\alpha}
\psi_{\beta} , \quad h \in Spin(1,p)\  \alpha , \beta = 1, 2, \dots
dim(D_{Spin(1,p)}) ,
$$
where the indices $\alpha$, $\beta$ enumerate the finite-dimensional
nonunitary or infinite-dimensional unitary $Spin(1,p)$ representation spaces.

The $\overline{SDiff}_{0}(p+1,R)$ spinor $\Psi$ transforms as in the previous
case. The $D_{\overline{Diff}_0(p+1,R)}$ representations can be reduced to a
direct sum of finite-dimensional or infinite-dimensional $D_{Spin(1,p)}$
representations. 

Owing to the fact that, in this case, both $\overline{SDiff}_{0}(p+1,R)$ and
$\overline{SL}(p+1,R)$ groups are represented nonlinearly over $Spin(1,p)$,
one has that both $SDiff_{0}(p+1,R)$ and $SL(p+1,R)$ groups are represented
nonlinearly over $SO(1,p)$ as well. Therefore, in this case there are no
usual, linearly transforming, $SL(p+1,R)$ tensorial quantities. Therefore,
this case seems to be of no importance for a spinning $p$-brane formulation
because it fails to provide for a group-theoretical formulation of the bosonic
theory sector.

\section{$\overline{SL}(p+1,R)$ representations construction}

We face now the problem of constructing the (unitary) infinite-dimensional
spinorial and tensorial representations of the $\overline{SL}(p+1,R)$ group.
The $\overline{SL}(p+1,R)$ group can be contracted (a la Wigner-In\"on\"u)
w.r.t. its $Spin(p+1)$ subgroup to yield the semidirect-product group ${\hat
T}\wedge Spin(p+1)$. ${\hat T}$ is an $\frac{1}{2}(p)(p+3)$ parameter Abelian
group generated by operators $U_{\hat i \hat j}$ $=$ $\lim_{\varepsilon\to 0}
(\varepsilon T_{\hat i \hat j})$, which form a $Spin(p+1)$ second rank
symmetric operator obeying the following commutation relations,
\begin{eqnarray}
&&[J_{\hat i\hat j},\ J_{\hat k\hat l}] = -i\delta_{\hat i\hat k}J_{\hat j
  \hat l} +i\delta_{\hat i \hat l}J_{\hat j \hat k}
+i\delta_{\hat j \hat k}J_{\hat i \hat l} -i\delta_{\hat j \hat l}j_{\hat i
  \hat k}, \nonumber \\ 
&&[ J_{\hat i\hat j},\ U_{\hat k\hat l}] = -i\delta_{\hat i\hat k}U_{\hat
  j\hat l} -i\delta_{\hat i\hat l}U_{\hat j\hat k}
+i\delta_{\hat j\hat k}U_{\hat i\hat l} +i\delta_{\hat j\hat l}U_{\hat i\hat
  k},  \\ 
&&[ U_{\hat i\hat j},\ U_{\hat k\hat l}] = 0 . \nonumber
\end{eqnarray}

An efficient way of constructing explicitly the $\overline{SL}(p+1,R)$
unitary infinite-dimen\-si\-onal representations is given by the
so called "decontraction" formula, which is an inverse of the
Wigner-In\"on\"u contraction. According to the decontraction formula,
the following operators
\begin{equation}
T_{\hat i\hat j} = r U_{\hat i\hat j} + \frac{i}{2\sqrt{U\cdot U}}
\left[ C_2(Spin(p+1)),\ U_{\hat i\hat j} \right],
\end{equation}
together with $J_{\hat i\hat j}$ form the $\overline{SL}(p+1,R)$ algebra. The
parameter $r$ is an arbitrary complex number, $r\in C$, and $C_2(Spin(p+1))$
is the $Spin(p+1)$ second-rank Casimir operator.

For the representation Hilbert space we take the homogeneous space of
$L^2$ functions of the maximal compact subgroup $Spin(p+1)$ parameters. The
$Spin(p+1)$ representation labels are given either by the Dynkin labels
$({\lambda}_1, {\lambda}_2,\dots , {\lambda}_q)$ or by the highest weight
vector which we denote by $\{ j\} = \{ j_1, j_2, \dots , j_q \}$, $q =
\left[\frac{p+1}{2}\right]$.
The $\overline{SL}(p+1,R)$ commutation relations are invariant w.r.t. an
automorphism defined by:
\begin{equation}
s(J) = +J, \quad s(T) = -T .
\end{equation}
This allows us to associate an '$s$-parity' to each $Spin(p+1)$ representation
contained in an $\overline{SL}(p+1,R)$ representation. In terms of the Dynkin
labels we find 
\begin{eqnarray}
s(D_2) &=& (-)^{{1 \over 2}({\lambda}_1 + {\lambda}_2 - \epsilon )},\nonumber
\\ s(D_{n\ge 3}) &=& (-)^{ {\lambda}_1 + {\lambda}_2 + \dots +
{\lambda}_{n-2} + {1\over 2}({\lambda}_n - {\lambda}_{n-1} - \epsilon
)} \\ 
s(B_1) &=& (-)^{ {1\over 2}({\lambda}_1 - \epsilon )} \nonumber \\
s(B_{n\ge 2}) &=& (-)^{ {\lambda}_1 + {\lambda}_2 + \dots +
{\lambda}_{n-1} + {1\over 2}({\lambda}_n - \epsilon )} \nonumber 
\end{eqnarray}
where $\epsilon = 0$ and $\epsilon =1$ for $\lambda$ even and odd,
respectively, and $D$ and $B$ refer to Cartan's Lie algebra notation.

The $s$-parity of the $\frac{1}{2}(p)(p+3)$-dimension representation
$(20\dots 0)$ $=$ $\Box\!\Box$ of $Spin(p+1)$ is: $s(\Box\!\Box )= +1$. A
basis of an $Spin(p+1)$ irreducible representation is provided by the
Gel'fand-Zetlin pattern characterized by the maximal weight vectors of the
subgroup chain $Spin(p+1)$ $\supset$ $Spin(p)$ $\supset$ $\cdots$ $\supset$
$Spin(2)$. We write the basic vectors as $\left| { \{ j \}   }\atop { \{m \}
  }\right>$, where $\{ j \}$ are the $Spin(p+1)$ group labels, and the
additional labels $\{ m \}$ corresponds to $Spin(p)$ $\supset$ $Spin(p-1)$
$\supset$ $\cdots$ $\supset$ $Spin(2)$ subgroup chain weight vectors. 

The Abelian group generators $\{ U\}$ $=$ $\{ U_{\{\mu \}}^{\{ \Box\!\Box \} }
\}$, $\{\mu \} = 1, 2,\dots , \frac{1}{2}(p)(p+3)$, can be, in the case of
multiplicity free representations, written in terms of the $Spin(p+1)$-Wigner
functions as follows, 
\begin{equation}
U_{\{ \mu \} }^{\{ \Box\!\Box \} } = D_{\{0\} \{\mu \}}^{\{\Box\!\Box \}} 
(\phi ) ,
\end{equation}
$\phi$ being $Spin(p+1)$ group parameters (e.g. Euler angles). 

It is now rather straightforward to determine explicitly the non-compact
operators matrix elements, which are given by the following expression:
\begin{eqnarray}
\left< \begin{array}{c} \{ j'\} \\ \{ m'\} \end{array} \right|  T_{\{
\mu \}}^{\{ \Box\!\Box \}}  \left| \begin{array}{c} \{ j\} \\ \{ m\}
\end{array} \right>  &=& \left( \begin{array}{ccc} \{ j'\} & \{
\Box\!\Box \} & \{ j\} \\ \{ m'\} & \{ \mu \} & \{ m\} \end{array}
\right)  \left< \{ j'\} \right|| T^{\{ \Box\!\Box \} } \left|| \{ j \}
\right>,   \\
\left< \{ j'\} \right|| T^{\{ \Box\!\Box \} } \left|| \{
j\} \right> &=& \sqrt{dim\{ j'\} dim\{ j\} }  \left\{ r + \frac{1}{2}
( C_2 (\{ j'\}) - C_2 (\{ j\} ) ) \right\}  \nonumber \\
&&\times \left(\begin{array}{ccc} \{ j'\} & \{ \Box\!\Box \} & \{ j\} \\
\{ 0\} & \{  0\} &  \{ 0\} \end{array}   \right).
\end{eqnarray}
$\pmatrix{\cdot&\cdot&\cdot\cr \cdot&\cdot&\cdot\cr}$ is the
appropriate $"3j"$ symbol for the $Spin(p+1)$ group. The (unitary)
infinite-dimensional representations of the $\overline{SL}(p+1,R)$ algebra are
given by these expressions of the non-compact generators together with the
well known representation expressions for the maximal compact $Spin(p+1)$
algebra generators. Finally, we apply the deunitarizing automorphism for a
correct physical interpretation. 

The very fact that the $\overline{SL}(p+1,R)$ generators are  constructed in
the basis of the maximal compact subgroup $Spin(p+1)$, i.e. in the Hilbert
space of square integrable functions, guarantees that they can be
exponentiated to the  corresponding $\overline{SL}(p+1,R)$ group
representations,
\begin{equation}
D_{\overline{SL}(p,R)}(e^{-i \zeta^{jk} T_{jk}} e^{-i \omega^{jk} J_{jk}} ) = 
e^{-i \zeta^{jk} D_{\overline{SL}(p,R)}(T_{jk})} 
e^{-i \omega^{jk} D_{\overline{SL}(p,R)}(J_{jk})} .
\end{equation}

In the case of the multiplicity free $\overline{SL}(p+1,R)$ representations,
each $Spin(p+1)$ sub-representation appears at most once and has the
same $s$-parity. This feature is especially useful for the task of reducing
infinite-dimensional spinorial and tensorial representations of the
$\overline{SL}(p+1,R)$ group to the corresponding $\overline{SL}(p,R)$
subgroup representations.

We present now just a few examples of the simplest $\overline{SL}(p+1,R)$
spinorial representations in terms of the corresponding $Spin(p+1)$ subgroup
representations. 
$$
p=2: \quad D_{\overline{SL}(3,R)} \supset D_{Spin(3)}^{2} \oplus
D_{Spin(3)}^{6} \oplus D_{Spin(3)}^{10} \oplus \dots , 
$$
$$
p=3: \quad D_{\overline{SL}(4,R)} \supset D_{Spin(4)}^{2} \oplus
D_{Spin(4)}^{6} \oplus D_{Spin(4)}^{12} \oplus \dots ,
$$
$$
p=4: \quad D_{\overline{SL}(5,R)} \supset D_{Spin(5)}^{4} \oplus
D_{Spin(5)}^{40} \oplus D_{Spin(5)}^{140} \oplus \dots ,
$$
$$
p=7: \quad D_{\overline{SL}(8,R)} \supset D_{Spin(8)}^{8} \oplus
D_{Spin(8)}^{56} \oplus D_{Spin(8)}^{224} \oplus \dots ,
$$
$$
p=9: \quad D_{\overline{SL}(10,R)} \supset D_{Spin(10)}^{16} \oplus
D_{Spin(10)}^{144} \oplus D_{Spin(10)}^{720} \oplus \dots ,
$$
where the $Spin(p+1)$ representation superscript denotes its dimensionality.

\section{The Spinning string case}

Let us finally address the question of a group-theoretical approach to 
construction of spinning $p$-brane infinite-dimensional Lie algebras that
generalize the Virasoro, and Neveu-Schwarz-Ramond algebras, and superalgebras,
respectively. 

Fradkin and Linetsky \cite{R19} proposed a method of constructing
infinite-dimensional Lie algebras (of the Virasoro type) by analytic
continuation of the finite classical algebras in the space of weight
diagrams. This method fails for $\overline{Diff}_0(p+1,R)$ and/or
$\overline{SL}(p+1,R)$ algebras, since in these cases there are no
finite-dimensional weight diagrams to be continued to an infinite system. 

We have explicitly constructed above the infinite-dimensional spinorial and
tensorial representations of the $\overline{SL}(p+1,R)$ group, over which the
full $p$-brane invariance $\overline{SDiff}_0 (p+1,R)$ is realized
nonlinearly. There are two relevant facts: (i) an action of the
$\overline{SDiff}_0 (p+1,R)$ generators leaves the $\overline{SL}(p+1,R)$
group representation space ${\cal V}_{\overline{SL}(p+1,R)}$ invariant, and
(ii) the $\overline{SDiff}_0 (p+1,R)$ generators $L_{(n)k}^{i_1i_2\dots
i_{n-1}}$, $n = 2, \dots \infty$ transform w.r.t. $\overline{SL}(p+1,R)$
subalgebra generators $L_{(2)k}^{i}$ as components of an irreducible tensor
operator. 

On the basis of these two facts, we propose the following procedure to
construct the infinite $p$-brane Lie algebras/superalgebras: 

(a) Introduce an infinite set of operators characterized by the
  $\overline{SL}(p+1,R)$ group representation labels, 

(b) Require these operators to have commutation relations with the
$L_{(2)k}^{i}$ generators as components of an irreducible tensor operator, and

(c) Demand that these operators satisfy mutually, as well as with the
  $\overline{SL}(p+1,R)$ generators, the (graded) Jacobi relations.

We demonstrate now this three steps procedure in the well known, $p =1$, case
of the spinning string Virasoro and Neveu-Schwarz-Ramond algebras.

\subsection{Irreducible representations of the $\overline{SL}(2,R)$ group}

The commutation relations of the $\overline{SL}(2,R)$ algebra $\{ J_0 ,
T_{\pm} = T_{1} \pm T_{2} \}$ read
$$
[J_{0},\ T_{\pm}]\ =\ \pm T_{\pm} \quad [T_{+},\ T_{-}]\ =\ -2J_{0}.
$$
According to the Iwasawa decomposition, $G = NAK$, where $N$, $A$, $K$ are
nilpotent, Abelian and maximal compact subgroups respectively. Any group
element $g \in G$ can be written as
$$
g = n(\nu )a(\lambda )k(\gamma ) =
\left(\matrix{1 & \nu \cr
              0 & 1   \cr}\right)
\left(
\matrix{\exp(\frac{\lambda}{2}) &           0             \cr
               0                & \exp(-\frac{\lambda}{2})\cr}
\right)
\left(
\matrix{\cos(\frac{\gamma}{2}) & -\sin(\frac{\gamma}{2})   \cr
        \sin(\frac{\gamma}{2}) &  \cos(\frac{\gamma}{2})   \cr}
\right).
$$

The differential forms of the group generators and the Casimir operator, in
terms of the above parameters, are 
$$
J_{0} = i\frac{\partial}{\partial\gamma}, \quad
T_{\pm} = e^{\mp i\gamma }
\Big( i\frac{\partial}{\partial\lambda}\
\mp \frac{\partial}{\partial\gamma}\Big); \quad\quad
C^{2} = \frac{\partial}{\partial\lambda}
\Big(\frac{\partial}{\partial\lambda}\ - 1\Big) .
$$
The generators matrix elements, in the $J_{0}$ eigenstate basis
$f_{m}(\gamma )$ $=$ $\langle\gamma\vert m\rangle$, $m=0,\pm
\frac{1}{2}, \dots $ ($\frac{\partial}{\partial\lambda} \rightarrow a$)
are as follows:
$$
J_{0}\vert m\rangle = m\vert m\rangle ,\quad  T_{\pm}\vert m\rangle =
i(a \pm m)\vert m\pm 1\rangle ;\quad C^{2}\vert m\rangle =
a(a-1)\vert m\rangle \quad \ \forall a.
$$

\subsection{Infinite bosonic algebra - Virasoro algebra}

Let $\{ E_m |  m = 0, \pm 1, \pm 2 \dots \}$ be an infinite set of operators,
such that $[ E , E ] \subset E$, which transform as components of
$\overline{SL}(2,R)$ irreducible tensor operator,
$$
[J_0 , E_m ] = m E_m , \quad [T_{\pm} , E_m ] = i(a \pm m) E_{m \pm 1} .
$$
The Jacobi relation for $(J_0 , E_m , E_n )$ implies
$$
[E_m , E_n ] = A_{m,n} E_{m+n} + C_m \delta_{m+n,0} ,
$$
while the Jacobi relation for $(T_{+} , E_m , E_n )$ implies
\begin{eqnarray*}
&&(a+m+n) A_{m,n} = (a+m) A_{m+1,n} + (a+n) A_{m,n+1} \\
&&(a+m) C_{m+1} + (a+n) C_{m} = 0 \quad m+n+1 = 0 .
\end{eqnarray*}
There is a solution of these relations for $a=-1$, and finally, we arrive at
the Virasoro algebra, i.e.
\begin{equation}
[E_m , E_n ] = (m-n) E_{m+n} + d m(m^2-1) \delta_{m+n+1,0} ,\quad d \in R .
\end{equation}

\subsection{Infinite super algebra - Neveu-Schwarz-Ramond superalgebra}

Let  $\{ E_m | m = 0, \pm 1, \pm 2 \dots \}$, and $\{ S_{\mu} | \mu = \pm
\frac{1}{2}, \pm \frac{3}{2}, \dots \}$, be infinite sets of operators, such
that  $[ E , E ] \subset E$, $[ E, S ] \subset S$ and $\{ S, S \} \subset E$,
which transform as components of $\overline{SL}(2,R)$ irreducible tensor
operators,
\begin{eqnarray*}
&[J_0 , E_m ]& = m E_m , \quad\quad
[T_{\pm} , E_m ] = i(a \pm m) E_{m \pm 1} , \\
&[J_0 , S_{\mu} ]& = \mu S_\mu , \quad\quad
[T_{\pm} , S_{\mu} ] = i(b \pm \mu ) S_{\mu \pm 1}
\end{eqnarray*}
The Jacoby relation for $(J_0 , S_{\mu} , S_{\nu} )$ implies
$$
\{ S_{\mu} , S_{\nu} \} = B_{\mu ,\nu} E_{m+n} + D_{\mu} \delta_{\mu +\nu ,0} ,
$$
while the Jacobi relation for $(T_{+} , S_{\mu} , S_{\nu})$ implies
\begin{eqnarray*}
&&(a+\mu + \nu) B_{\mu ,\nu} =
(b+\mu ) B_{\mu +1,\nu} + (b+\nu ) B_{\mu ,\nu +1} \\
&&(b+\mu ) D_{\mu +1} + (b+\nu ) D_{\mu} = 0 \quad \mu +\nu +1 = 0 .
\end{eqnarray*}
There is a solution of these equations for $b = -\frac{1}{2}$, 
$$
\{ S_{\mu} , S_{\nu} \} = 2 E_{\mu + \nu} + d' (\mu^2 - \frac{1}{4})
\delta_{\mu + \nu , 0} .
$$
The Jacobi relation for $(J_{0} , E_{m} , S_{\mu})$ implies
$$
[E_{m} , S_{\mu} ] = F_{m,\mu} S_{m+\mu} ,
$$
while the Jacobi relation for $(T_{+} , E_{m} , S_{\mu})$ implies
$$
(b + m + \mu ) F_{m, \mu} = (a + m) F_{m+1,\mu} + (b + \mu )F_{m,\mu +1} ,
$$
and for $a = -1$, $b = -\frac{1}{2}$ one has
$$
F_{m,\mu } = \left( \frac{m}{2} - \mu \right) F .
$$
The Jacobi relation for $(E_{m} , E_{n} , S_{\mu })$ implies
$$
\left(\frac{n}{2}-\mu\right) \left(\frac{m}{2}-n-\mu \right) F^2
= (m-n) \left(\frac{m}{2}+\frac{n}{2}-\mu\right) F
+ \left(\frac{m}{2}-\mu \right) \left(\frac{n}{2}-m-\mu\right) F^2
$$
For $F=1$ one has
$$
[ E_{m}, S_{\mu} ] = \left(\frac{m}{2}-\mu\right) S_{m+\mu} .
$$
The Jacobi relation for $(S_{\mu} , E_{m} , S_{\nu})$ implies
$$
d' \left( \frac{m}{2} -\nu \right) \left(\mu^2 -\frac{1}{4} \right)
= d' \left(\mu -\frac{m}{2} \right) \left( (m+\mu )^2 - \frac{1}{4} \right)
+ 2dm (m^2 -1) ,
$$
i.e. $d' = 4d$

Finally, we obtain the Neveu-Schwarz-Ramond superalgebra:

\begin{eqnarray}
&[E_m , E_n ]& = (m-n) E_{m+n} + d m(m^2-1) \delta_{m+n+1,0} ,\nonumber \\
&[E_m , S_{\mu} ]& = \left(\frac{m}{2}-\mu \right) S_{m+\mu}, \\
&\{ S_{\mu} , S_{\nu} \}& = 2 E_{\mu +\nu} +
4 d \left(\mu^2 -\frac{1}{4} \right) \delta_{\mu +\nu , 0} \quad d \in R . 
\nonumber 
\end{eqnarray}

\section*{Acknowledgments}

This work was supported in part by MS RS Project-141036.

\end{document}